\documentclass[fleqn,usenatbib]{mnras}

\usepackage[T1]{fontenc}

\DeclareRobustCommand{\VAN}[3]{#2}
\let\VANthebibliography\thebibliography
\def\thebibliography{\DeclareRobustCommand{\VAN}[3]{##3}\VANthebibliography}

\newcommand{\Msun}{\,{\rm M_\odot}}
\newcommand{\fedd}{\,{f_{\rm Edd}}}
\newcommand{\Mblack}{M_\bullet}

\usepackage{graphicx}	
\usepackage{amsmath}	
\usepackage{amssymb}	

\usepackage{tipa}

\title[The Search for the Farthest Quasar]{The Search for the Farthest Quasar: Consequences for Black Hole Growth and Seed Models}

\author[F. Pacucci \& A. Loeb]{
Fabio Pacucci$^{1,2}$\thanks{fabio.pacucci@cfa.harvard.edu},
Abraham Loeb$^{1,2}$
\\
$^{1}$Center for Astrophysics $\vert$ Harvard \& Smithsonian, Cambridge, MA 02138, USA\\
$^{2}$Black Hole Initiative, Harvard University, Cambridge, MA 02138, USA\\
}

\date{\today}

\pubyear{2021}

\begin{document}
\label{firstpage}
\pagerange{\pageref{firstpage}--\pageref{lastpage}}
\maketitle

\begin{abstract}
The quest for high-redshift quasars has led to a series of record-breaking sources, with the current record holder at $z=7.642$. Here, we show how future detections of $z>8$ quasars impact the constraints on the parameters for black hole growth and seed models. Using broad flat priors on the growth parameters (Eddington ratio $\fedd$, duty cycle ${\cal D}$, seed mass $M_{\rm \bullet, seed}$ and radiative efficiency $\epsilon$), we show that the large uncertainties in their determination decrease by a factor $\sim 5$ when a quasar's detection redshift goes from $z=9$ to $z=12$. In this high-redshift regime, $\epsilon$ tends to the lowest value allowed, and the distribution for $M_{\rm \bullet, seed}$ peaks well inside the heavy seed domain. Remarkably, two quasars detected at $z > 7$ with low accretion rates (J1243+0100 and J0313--1806) already tighten the available parameter space, requiring $M_{\rm \bullet, seed} > 10^{3.5} \Msun$ and $\epsilon < 0.1$. The radiative efficiency is a crucial unknown, with factor $\sim 2$ changes able to modify the predicted mass by $\sim 3$ orders of magnitude already at $z\sim 9$. The competing roles of inefficient accretion (decreasing $\epsilon$) and black hole spin-up (increasing $\epsilon$) significantly impact growth models.
Finally, we suggest that yields currently predicted by upcoming quasar surveys (e.g., Euclid) will be instrumental for determining the most-likely seed mass regime. For example, assuming thin-disk accretion, a detection of a quasar with $\Mblack \sim 10^{10} \Msun$ by $z\sim 9-10$ would exclude the entire parameter space available for light seeds and dramatically reduce the one for heavy seeds.
\end{abstract}

\begin{keywords}
black hole physics -- methods: statistical -- galaxies: active -- quasars: general -- early Universe -- surveys
\end{keywords}

\section{Introduction} \label{sec:intro}
The last half century in the search for high-$z$ quasars has led to a chain of detections, constantly breaking the record for the farthest one (e.g., \citealt{Fan_2001, Fan_2003, Mortlock_2011, Wu_2015, Banados_2018, Yang_2020, Wang_2021}). Quasars are powered by super-massive black holes (SMBHs) with mass $\gtrsim 10^9 \Msun$, typically accreting at rates comparable to the Eddington limit, i.e. the rate at which the outward radiation acceleration balances the inward gravitational acceleration. 
Currently, the record for the highest redshift quasar is held by J0313--1806, a source at $z=7.642$ powered by a SMBH of $(1.6 \pm 0.4) \times 10^9 \, \Msun$ \citep{Wang_2021}. Based on its redshift, J0313--1806 is shining a mere $\sim 680$ Myr after the Big Bang. The detection of farther quasars is putting a significant strain on the standard theory of black hole growth (see, e.g., \citealt{Woods_2019, Inayoshi_review_2019} for recent reviews), as it is becoming increasingly challenging to build up $\gtrsim 10^9 \Msun$ SMBHs in time to match high-$z$ observations. This problem with high-$z$ quasars was explored already by \cite{Turner_1991}, when the highest redshift quasars were detected merely at $4<z<5$. More forcefully, the growing issue was pointed out by \cite{Haiman_Loeb_2001} with the detections of the first $z \sim 6$ quasars \citep{Fan_2001}.

The problem of how to form these SMBHs quickly enough to match these observations remains open. The first population of black holes, commonly referred to as black hole seeds, likely formed around the cosmic time when also the first population of stars (Pop III) were born, at $z \sim 20-30$, or $\sim 200$ Myr after the Big Bang \citep{BL01}. This would leave $\sim 500$ Myr of cosmic time from the formation of the first seeds to the observation of the currently farthest, fully-fledged SMBHs \citep{Loeb_2013}.
Depending on their initial mass, seeds are commonly categorized into light seeds and heavy seeds, with typical mass $\Mblack \sim 10^2 \Msun$ and $\Mblack \sim 10^5 \Msun$, respectively (see, e.g., \citealt{Inayoshi_review_2019} and references therein).

Forming a $\sim 10^9 \Msun$ SMBH by $z \sim 7$ from a light seed is very challenging, unless periods of  super-Eddington or hyper-Eddington accretion rates can be sustained by the growing black hole (e.g., \citealt{Begelman_1978, Wyithe_Loeb_2012, Inayoshi_2016, Begelman_Volonteri_2017, Alexander_2014, Pacucci_2017, Natarajan_2021}). Note that while super-Eddington accretion rates are typically of the same order of magnitude of the Eddington rate, hyper-Eddington episodes can reach hundreds or even thousands times larger values, due to photon trapping conditions quenching the effect of radiation pressure on the infalling gas \citep{Wyithe_Loeb_2012, Inayoshi_2016, Begelman_Volonteri_2017}.

Formation mechanisms for heavy seeds started to being developed at the turn of the century in order to jump-start the growth process and possibly ease the challenges of forming quasars by $z \sim 6$. Typically, they are formed by one of the following mechanisms.
First, a monolithic collapse of a pristine gas cloud in atomic cooling halos, where the cooling action of hydrogen molecules is quenched by some irradiation/heating mechanisms \citep{Loeb_Rasio_1994, Bromm_Loeb_2003, Lodato_Natarajan_2006}, possibly undergoing to a short-lived phase of super-massive star \citep{Loeb_Rasio_1994, Hosokawa_2013}. These seeds are typically in the intermediate mass range $10^{5-6} \Msun$.
Second, runaway collisions and mergers of either Pop III stars \citep{PZ_2002, Katz_2015, Boekholt_2018} or black holes in a dense gaseous environment \citep{Davies_2011, Lupi_2016}. These seeds are typically in the mass range $10^{3-4} \Msun$.

As future telescopes will enable deeper surveys of high-$z$ quasars (e.g., Euclid, the Roman Space Telescope, Lynx, the James Webb Space Telescope), it is a prime time to discuss the implications of future detections of quasars at $z > 8$ on growth and seed models of black holes. Note that the first data release of Euclid is predicted to yield between 13 and 25 quasars in the redshift range $7<z<9$, depending on the rate of decrease of the spatial density of quasars as a function of redshift (e.g., \citealt{Euclid_2019, Fan_2019BAAS}). This decrease is estimated from observations at lower redshift, typically $4 \lesssim z \lesssim 6$ \citep{McGreer_2013, Jiang_2016}, and then the linear extrapolation at $z \gtrsim 7$ is used to compute yields \citep{Euclid_2019}. Depending on the steepness of the density decrease, the farthest $\Mblack \gtrsim 10^9 \Msun$ quasar in the observable Universe is currently estimated to be found between $z=9$ and $z=12$ \citep{Ben-Ami_2018, Fan_2019BAAS}.

The highest redshift at which we can observe a $\Mblack \gtrsim 10^9$ SMBH is crucially important to constrain the fundamental parameters of black hole growth, as well as the typical seed mass required. In this paper, we discuss the implications of the detection redshift of the ``farthest quasar in the Universe''.

Before delving into the description of the theoretical model in \S \ref{sec:theory} and the results in \S \ref{sec:results}, a discussion of what we mean by ``farthest quasar'' is warranted. Of course, black holes of small mass and accreting at low rates should exist at redshifts $z \gg 7$, possibly at $z\sim 20-30$ \citep{BL_2000}, even if they will not observable in the foreseeable future. These smaller, slowly accreting black holes are not necessarily to be considered ``quasars''.
Previous studies (e.g., \citealt{Fan_2019BAAS, Wang_2019_density}) predict that the highest-redshift quasar observable should be at $z\sim 9$, defining quasar as a SMBH with mass $\Mblack > 10^9 \Msun$. This definition is supported by current observations, as we are only able to observe SMBHs on that mass scale at $z \gtrsim 7$. Our study, though, deals with future observations of SMBHs at $z > 7$, which may probe mass scales much lower than $\sim 10^9 \Msun$. Hence, in this study, ``farthest quasar'' is intended, more broadly, as the highest-redshift SMBH with mass $\Mblack > 10^6 \Msun$ and accreting at rates larger than $10\%$ of their Eddington rate, based on up-to-date accretion rate distributions of high-$z$ quasars \citep{Fan_2019}.

\section{THEORY} 
\label{sec:theory}
First we describe the simple growth model used along with the Monte Carlo simulations performed.

\subsection{Black Hole Growth Model} 
\label{subsec:growth_model
}
We begin by defining the most important parameters of the growth model. A black hole seed of initial mass $M_{\rm \bullet, seed}$ is seeded at $z=25$, corresponding to a cosmic time $t_{\rm seed} = 130$ Myr. We calculate the mass $\Mblack$ at cosmic time $t$ with the well-known formula:
\begin{equation}
    \Mblack(t) = M_{\rm \bullet, seed} \exp{ \left[ \fedd {\cal D} \frac{1-\epsilon}{\epsilon} \frac{\Delta t}{t_{\rm Edd} } \right] } \, .
    \label{eq:growth}
\end{equation}

The accretion time (also known as the lifetime of the quasar, see, e.g., \citealt{Eilers_2021}) is defined as $\Delta t = t-t_{\rm seed}$, while the Eddington time $t_{\rm Edd} \approx 450 \, \mathrm{Myr}$ is a combination of constants.
The Eddington ratio $\fedd$ is defined as the ratio between the actual mean accretion rate and the mean Eddington accretion rate over the time interval $\Delta t$. As the black hole mass increases, also its instantaneous Eddington accretion rate increases as $\dot{M}_{\rm Edd} \propto \Mblack$. The duty cycle ${\cal D}$ is the fraction of the time $\Delta t$ spent accreting. 
Note that in Eq. (\ref{eq:growth}) the quantities $\fedd$ and ${\cal D}$ are degenerate: the same final mass can be obtained by appropriately rescaling the two growth parameters by a factor.
Finally, $\epsilon$ is the mean radiative efficiency factor over the time $\Delta t$. The radiative efficiency factor determines the fraction of the accretion energy per unit time, $\dot{M}c^2$, that is converted into radiative luminosity: $L = \epsilon \dot{M}c^2$. For thin disk accretion, the value of $\epsilon$ varies between $5.7\%$ for non-rotating black holes to $32\%$ for maximally rotating black holes (see \citealt{Bardeen_1970} and the discussion in, e.g., \citealt{Fabian_2019, Pacucci_2020}).

\subsection{Monte Carlo Simulations} 
\label{subsec:MC}
We perform Monte Carlo simulations of the growth of black holes using Eq. \ref{eq:growth} as the target function. We fix the seeding redshift $z=25$ corresponding to a $t_{\rm seed} = 130$ Myr.
According to the latest high-$z$ surveys \citep{McGreer_2013, Jiang_2016, Wang_2019_density, Fan_2019BAAS}, the space density of quasars is rapidly declining above $z\sim 6$ and the farthest quasar in the Universe with $\Mblack \gtrsim 10^9 \Msun$ should be observable between $z=9$ and $z=12$ (see, e.g., Fig. 1 in \citealt{Fan_2019BAAS} and the discussion in \citealt{Euclid_2019}). Of course, these estimates are based on extrapolations from quasar luminosity functions at lower redshifts, and may be a poor description of the earlier Universe. Here, in line with current predictions, we consider detection redshifts for quasars between $z=9$ and $z=12$. Also, consider that with a seeding at $z=25$, an observation at $z=9$ would correspond to a time $\Delta t \sim  450$ Myr, equal to the e-folding Eddington time. 

We consider a growth model for the Monte Carlo simulation described by four parameters: $\left[ \fedd, {\cal D}, \epsilon, M_{\rm \bullet, seed} \right]$. Note that the seed mass is by definition a constant over $\Delta t$ (i.e., the black hole is seeded only once). ${\cal D}$ is the fraction of $\Delta t$ spent accreting, while ${\fedd}$ and $\epsilon$ can vary during $\Delta t$. Hence, the values of $\fedd$ and $\epsilon$ should be considered as averages over the total time $\Delta t$ between seeding and detection.

We divide the redshift range $9<z<12$ into $10$ bins and for each detection redshift $z_{\rm det}$ we study the distribution of $10^3$ realizations of the growth model that allow the formation of a $10^9 \Msun$ black hole in a time equal or shorter than the cosmic time between the seeding redshift $z=25$ and $z_{\rm det}$. We then save the combinations of ``successful'' growth parameters $\left[ \fedd, {\cal D}, \epsilon, M_{\rm \bullet, seed} \right]$. The mass $10^9 \Msun$ was chosen as representative of typical masses of quasars found at $z > 6$ \citep{Inayoshi_review_2019}.

In our baseline analysis, we consider the following priors on the four parameters.
The duty cycle has a flat prior: ${\cal D} \in [0,1]$.
The radiative efficiency factor has a flat prior within its physically meaningful range: ${\epsilon} \in [0.057,0.32]$. Light seed masses have a flat prior in the logarithm of the mass range: $\mathrm{Log_{10}} {M_{\rm \bullet, L}} \in [0,3]$, while heavy seed masses have a flat prior in the logarithm of the mass range: $\mathrm{Log_{10}}{M_{\rm \bullet, H}} \in [3,6]$ (e.g., \citealt{Inayoshi_review_2019}). Note that these ranges are very generous, as typical masses for light and heavy seeds are $\sim 10^2 \Msun$ and $10^5 \Msun$, respectively.
Finally, the Eddington ratio prior distribution is the most critical. While episodes of super-Eddington or even hyper-Eddington accretion likely occurred \citep{Inayoshi_2016, Begelman_Volonteri_2017} at $z > 6$, with Eddington ratios of the order of even $100-1000$, choosing a prior distribution of $\fedd$ that includes such extreme values would be unwise for two reasons: (i) hyper-Eddington accretion episodes are short-lived, with typical duration times of $\sim 0.1$ Myr \citep{Inayoshi_2016}, hence the mean would be much lower; (ii) choosing an upper limit on the prior distribution that is too large would artificially push the final distribution to the large values, in order to form the SMBHs in time.
Hence, we choose a flat prior $\fedd \in [0,1]$ as well. This choice is supported by observations: high-$z$ quasar surveys have found typical distributions of Eddington ratios distributed as a lognormal with a peak at $\fedd \sim 1.07$ and a standard deviation of $0.28$ dex \citep{Willott_2010}. This distribution of Eddington ratios is, by definition, at detection: this does not imply that the SMBH was accreting at similarly large accretion rates for its entire history between seeding and observation. Limitations and caveats of this approach are discussed more broadly in \S \ref{sec:disc_concl}.

The Monte Carlo simulations and Eq. (\ref{eq:growth}) are used in two separate settings, whose results are described in turn in \S \ref{sec:parameters} and \S \ref{sec:max_mass}.
In the first kind of analysis, we fix a final mass at the detection redshift $z_{\rm det}$ and investigate the probability distributions of parameters that can reproduce that observation, along with the accuracy with which they can be determined. This analysis is also applied in \S \ref{sec:from_data} to discuss constraints from current detections.
In the second kind of analysis, we leave the final mass at the detection redshift as a free parameter, and investigate the maximum mass achievable as a function of $z$, starting from different models of black hole seeds and radiative efficiencies.

\section{RESULTS} 
\label{sec:results}
Next we present the results of our study, concerning the determination of the growth parameters with redshift (\S \ref{sec:parameters}), its application to current data (\S \ref{sec:from_data}), and the maximum mass achievable in different seeding models and accretion conditions (\S \ref{sec:max_mass}).

\subsection{Determination of Growth Parameters} 

\begin{figure}
\includegraphics[angle=0,width=0.49\textwidth]{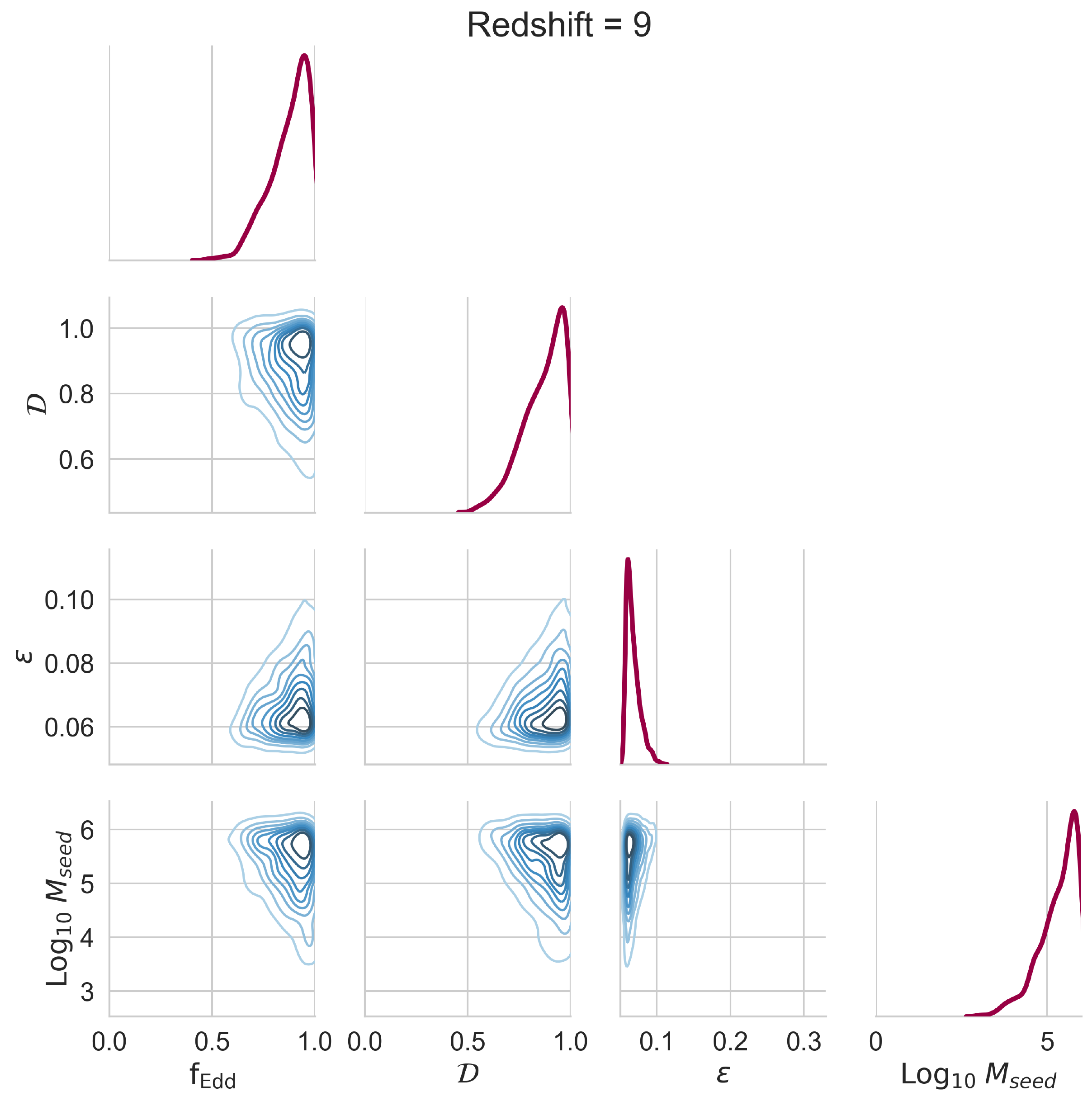}
\includegraphics[angle=0,width=0.49\textwidth]{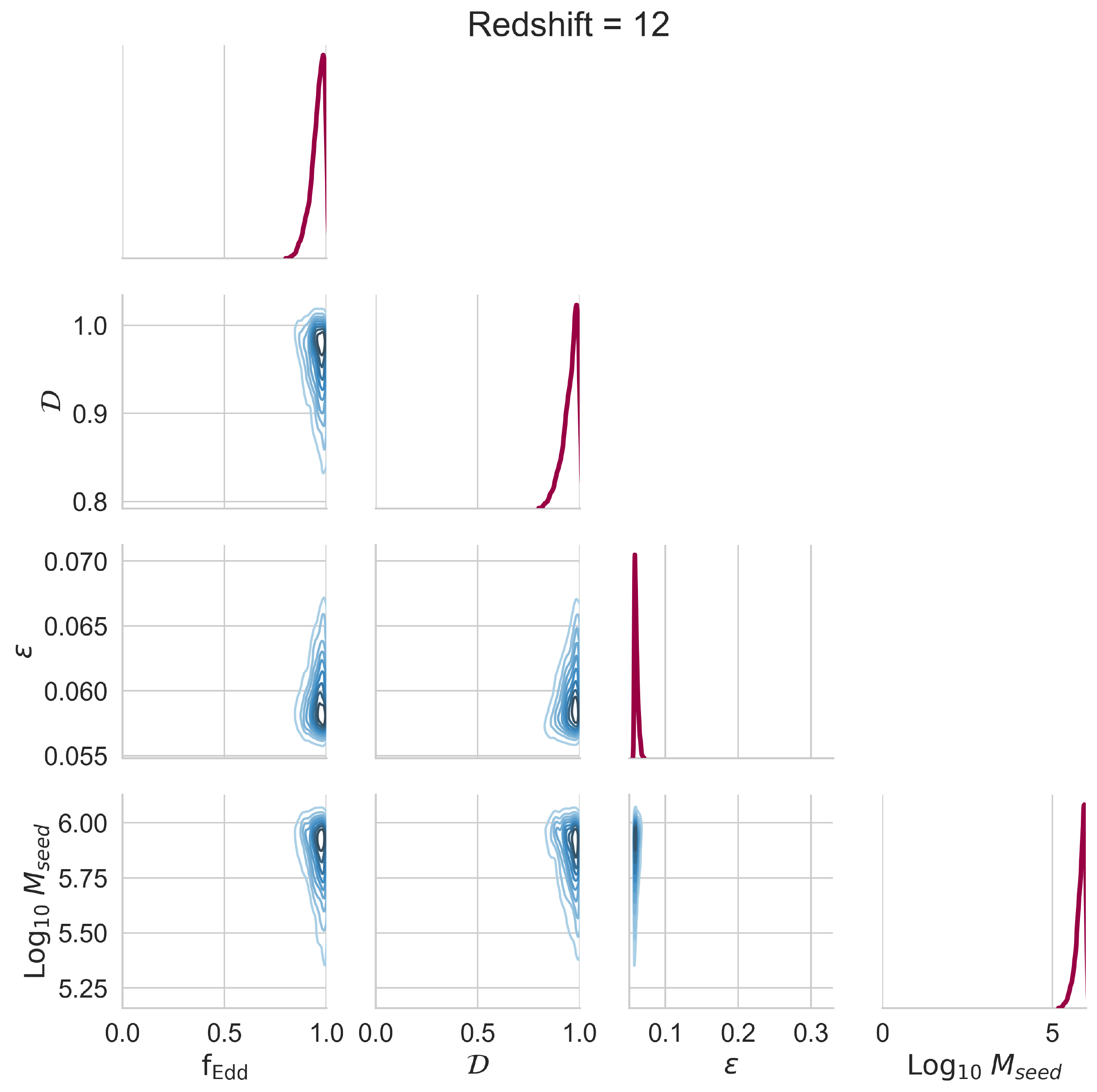}
\caption{Statistical distributions (2D joint in the lower half and 1D marginalized in the diagonal) of the sets of $\left[ {f_{\rm edd}}, {\cal D}, \epsilon, M_{\rm \bullet, seed} \right]$ that can produce at least a $10^9 \Msun$ SMBH by the detection redshift $z_{\rm det}$, with $z_{\rm det} = 9, 12$.} 
\label{fig:Cornerplot}
\end{figure}

\label{sec:parameters}

\begin{table}
\centering
\begin{tabular}{cccc}
\hline
\multicolumn{1}{l}{}                                              & $z=6$             & $z=9$             & $z=12$            \\ \hline
\multicolumn{1}{c|}{$\fedd$}                                & $0.81 \pm 0.15$ & $0.88 \pm 0.10$ & $0.96 \pm 0.03$ \\ \hline
\multicolumn{1}{c|}{${\cal D}$}                                   & $0.81 \pm 0.14$ & $0.88 \pm 0.1$ & $0.96 \pm 0.03$ \\ \hline
\multicolumn{1}{c|}{$\epsilon$}                                   & $0.08 \pm 0.02$ & $0.07 \pm 0.01$ & $0.060 \pm 0.002$ \\ \hline
\multicolumn{1}{c|}{$M_{\rm \bullet, seed}$} & $4.49 \pm 1.33$   & $5.34 \pm 0.57$  & $5.84 \pm 0.13$  \\ \hline
\label{table:1}
\end{tabular}
\caption{Statistical descriptors of the distributions shown in Figs. \ref{fig:Cornerplot} and, as a comparison, for $z_{\rm det}=6$. The values are reported as mean $\pm$ standard deviation. The value of $M_{\rm \bullet, seed}$ is expressed as $\mathrm{Log_{10}}$.}
\end{table}

The parameter space $\left[ \fedd, {\cal D}, \epsilon, M_{\rm \bullet, seed} \right]$ that allows the formation of a $\Mblack \geq 10^9 \Msun$ black hole at a given detection redshift $z_{\rm det}$ is shown in Fig. \ref{fig:Cornerplot}, for redshifts $z=9, 12$, the extreme values of the current redshift range where the farthest $\Mblack \gtrsim 10^9 \Msun$ detectable quasar should be located \citep{Fan_2019BAAS, Euclid_2019}.

A statistical description of the distribution, for $z_{\rm det} = 9$ and $z_{\rm det} = 12$, as shown in Fig. \ref{fig:Cornerplot}, and at $z_{\rm det}=6$ for comparison, is provided in Table \ref{table:1}. Each distribution is described with a set of mean $\pm$ standard deviation. 

From these results it is apparent that the uncertainties in the determination of the growth parameters are rapidly shrinking with increasing detection redshift. In particular, the standard deviations are typically reduced by a factor $10$ in shifting from $z_{\rm det} = 6$ to $z_{\rm det} = 12$, and by a factor $\sim 5$ merely from $z_{\rm det} = 9$ to $z_{\rm det} = 12$. The Eddington ratio and duty cycle are fully degenerate and tend to values $\sim 1$ at the highest detection redshifts. The radiative efficiency tends to values close to the limit for thin-disk, non rotating black holes, i.e. $\epsilon = 0.057$, while the seed mass very significantly tends to the heavy seed mass range, with $\mathrm{Log_{10}} \, M_{\rm \bullet, seed} = 5.84 \pm 0.13$ at the highest detection redshift.

It is important to note that this simple model is based on very generous flat priors and does not rely on any cross dependency between the parameters. For example, a determination that high-$z$ quasars have large spins, possibly generated by stable accretion via a disk, would dramatically reduce the parameter space of the radiative efficiency, leading to values closer to $\epsilon \sim 0.32$. This would, in turn, shift the acceptable ranges for other parameters significantly. A discussion of some of these cross dependencies in provided in \S \ref{sec:disc_concl}.

\begin{figure}
\includegraphics[angle=0,width=0.49\textwidth]{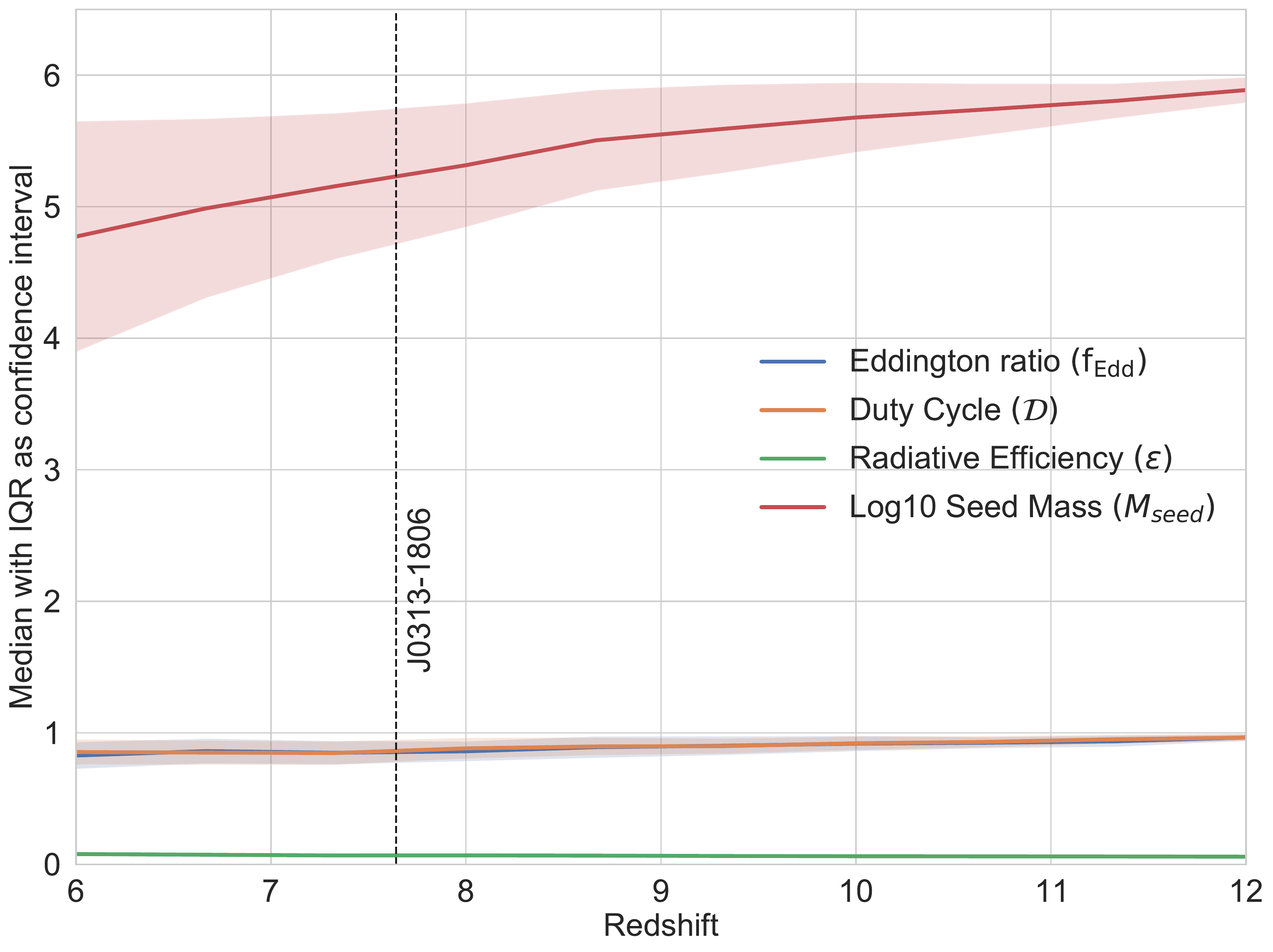}
\includegraphics[angle=0,width=0.49\textwidth]{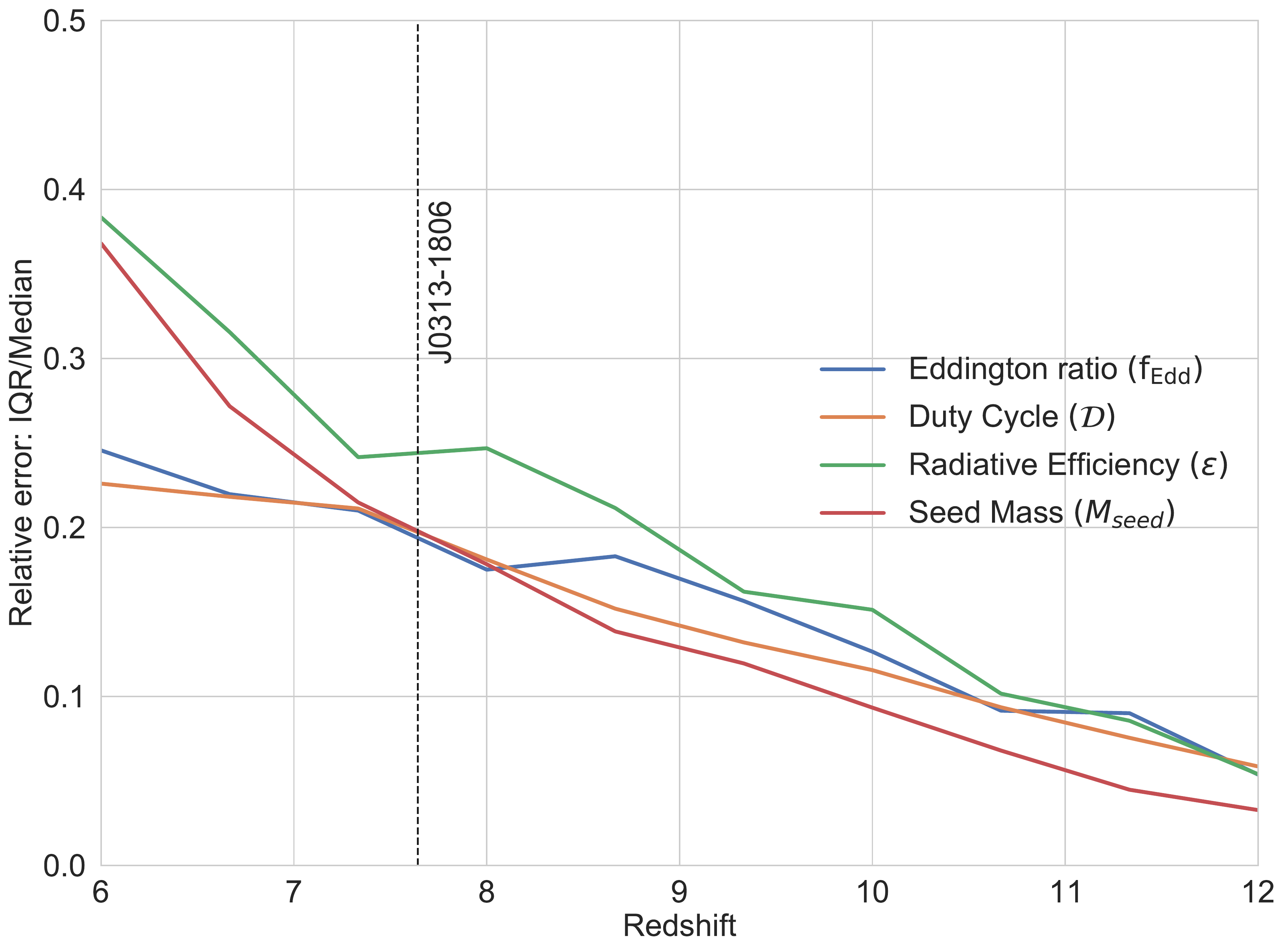}
\caption{\textbf{Top panel:} Typical values of the growth parameters, along with their uncertainties shown as an inter-quartile range (IQR). \textbf{Bottom panel:} Relative errors (IQR/median value) in the estimates of the growth parameter. Both are reported as a function of the detection redshift.} 
\label{fig:parameters_accuracy}
\end{figure}

The typical values of the parameters, along with their uncertainties reported as an inter-quartile range (IQR), are shown in the left panel of Fig. \ref{fig:parameters_accuracy}. The IQR was chosen instead of the standard deviation because some of the distributions are skewed around the median.

The relative errors (IQR/median value) in the estimates of the growth parameter, as a function of the detection redshift, are shown in the top panel of Fig. \ref{fig:parameters_accuracy}. The relative errors decrease in a similar manner by a factor $\sim 5$ from $z_{\rm dec} = 6$ to $z_{\rm dec} = 12$. This analysis shows that, even with a blind and unconstrained search of the growth parameter space, the detection of quasars between $z=9$ and $z=12$ would significantly increase the accuracy in our estimates of such parameters.

\subsection{Constraints from Current Detections} 
\label{sec:from_data}

\begin{figure}
\includegraphics[angle=0,width=0.49\textwidth]{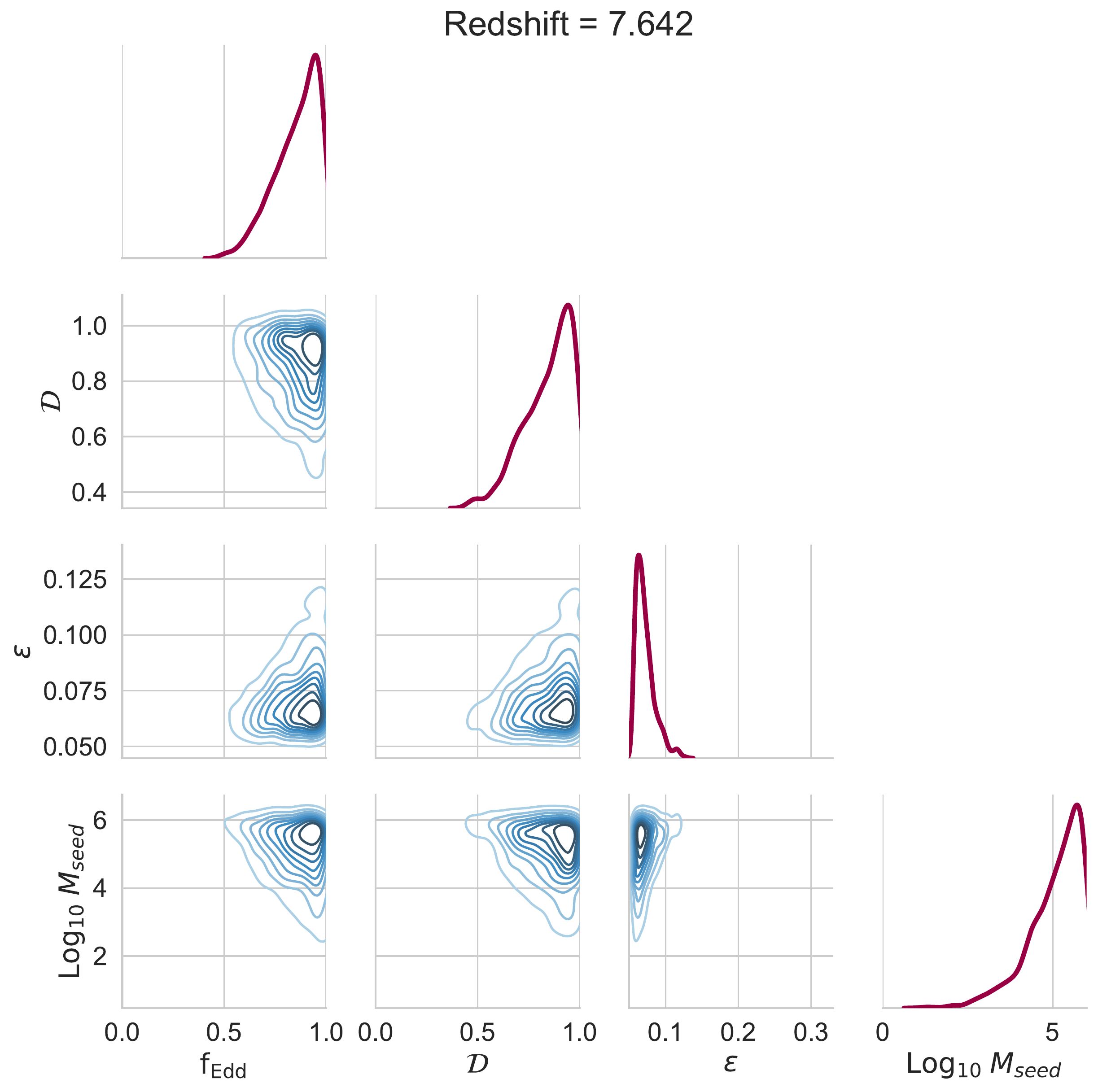}
\caption{As in Fig. \ref{fig:Cornerplot}, but for the currently highest-redshift quasar detected, with $\Mblack = 1.6\times 10^9 \Msun$ and $z_{\rm det} = 7.642$. The detection redshift is already sufficiently high to discard some areas of the parameter space.} 
\label{fig:Cornerplot_current}
\end{figure}

In Fig. \ref{fig:Cornerplot_current} we provide the same statistical analysis as in Fig. \ref{fig:Cornerplot}, but for the current farthest quasar detected \citep{Wang_2021} at $z=7.642$ and with mass $\Mblack = 1.6 \times 10^9 \Msun$. While the detection redshift is still low to significantly tighten the allowed parameter space, it is already possible to draw some conclusions. For example, Fig. \ref{fig:Cornerplot_current} shows that seed masses $M_{\rm \bullet, seed} \lesssim 10^{3}$ are disfavored. Note that large values of the accretion efficiency $\epsilon > 0.1$ are already disfavored, while intermediate values of $\fedd \sim 0.5$ and ${\cal D} \sim 0.5$ are still acceptable, although statistically disfavored.

\begin{figure}
\includegraphics[angle=0,width=0.49\textwidth]{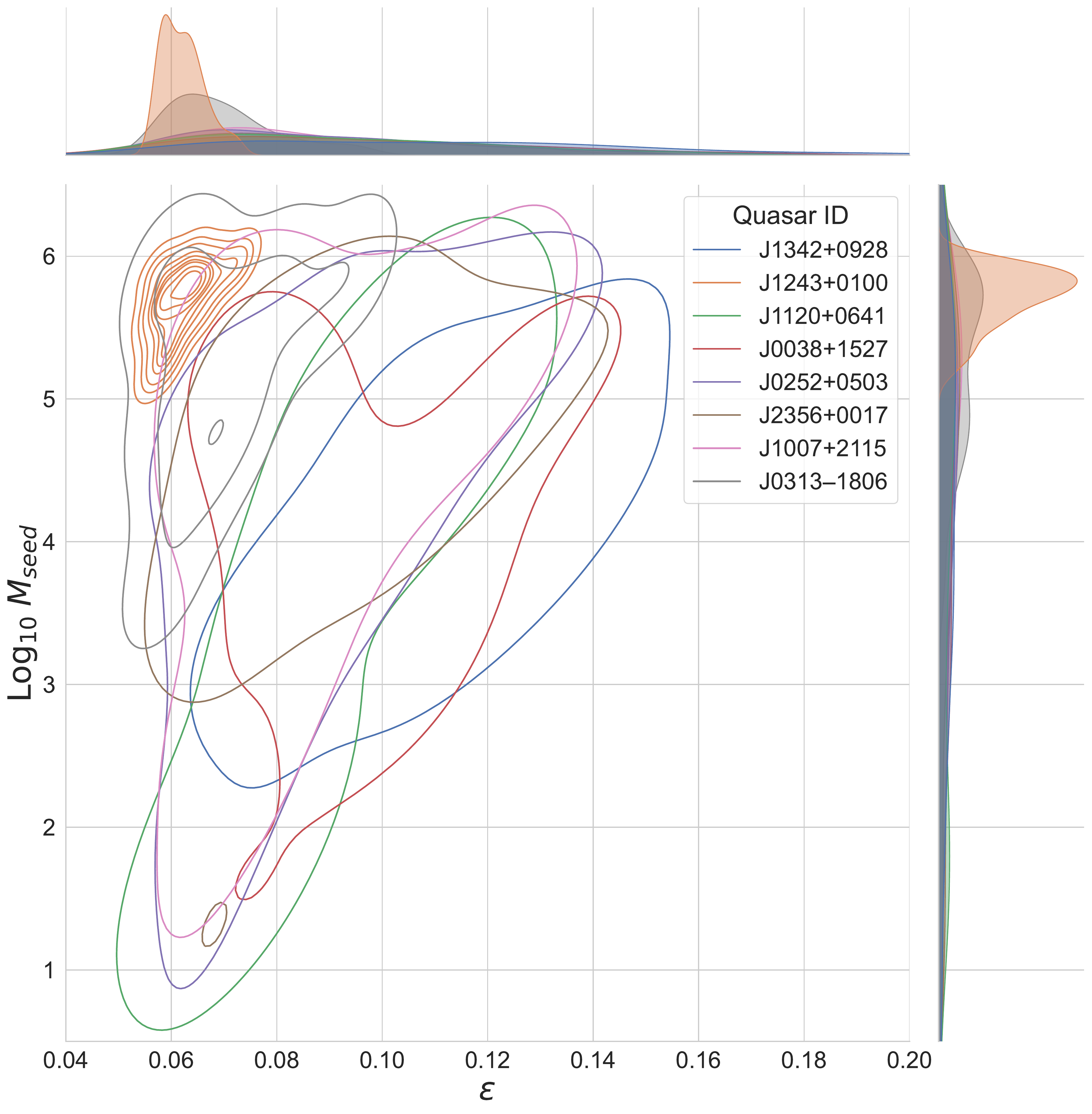}
\caption{Kernel density estimate (KDE) and marginal distributions for $\epsilon$ and $M_{\rm \bullet, seed}$ for eight quasars detected at $z > 7$ and with estimated values of the $\fedd$, which are used as priors. This plot assumes, for simplicity of representation, ${\cal D} = 1$.}
\label{fig:quasars}
\end{figure}

Several quasar detections come with an estimate of the Eddington ratio $\fedd$, which can be used as a prior in our model to gain further insights into the remaining parameters. Assuming for simplicity of representation ${\cal D} = 1$, in Fig. \ref{fig:quasars} we show the results of this analysis. We display the kernel density estimate (KDE) and marginal distributions of $\epsilon$ and $M_{\rm \bullet, seed}$ for eight quasars detected at $z > 7$ \citep{Mortlock_2011, Wang_2018, Banados_2018, Matsuoka_2019a, Matsuoka_2019, Yang_2019, Yang_2020, Wang_2021}. Note that this same dataset was used in \cite{Vagnozzi_2021} to estimate the ages of the oldest objects in the Universe.

As expected, most of these detections are still at low redshift and cannot tightly constrain the parameter space. For example, in some cases the seed mass ranges from $\sim 10$ to $\sim 10^6 \Msun$, and the radiative efficiency reaches values up to $\sim 0.15$.

Interestingly, the two quasars J1243+0100 and J0313--1806 show much tighter parameter spaces, because their estimated Eddington ratios are $\fedd = 0.34 \pm 0.2$ \citep{Matsuoka_2019a} and $\fedd = 0.67 \pm 0.14$ \citep{Wang_2021}, respectively. These quasars are accreting at lower Eddington rates (i.e., at a slower pace), and this significantly tightens the allowed parameter space, with $M_{\rm \bullet, seed} > 10^5 \Msun$ for J1243+0100 and $M_{\rm \bullet, seed} > 10^{3.5} \Msun$ for J0313--1806, while also requiring lower radiative efficiencies $\epsilon \lesssim 0.1$. Remarkably, the detection of additional quasars with low values of $\fedd$ will also tighten the available parameter space, without requiring the very high redshifts necessary for sources accreting at $\fedd \sim 1$.

\subsection{Maximum Black Hole Mass: Dependence on Seed Models and Radiative Efficiency} 
\label{sec:max_mass}
\begin{figure}
\includegraphics[angle=0,width=0.49\textwidth]{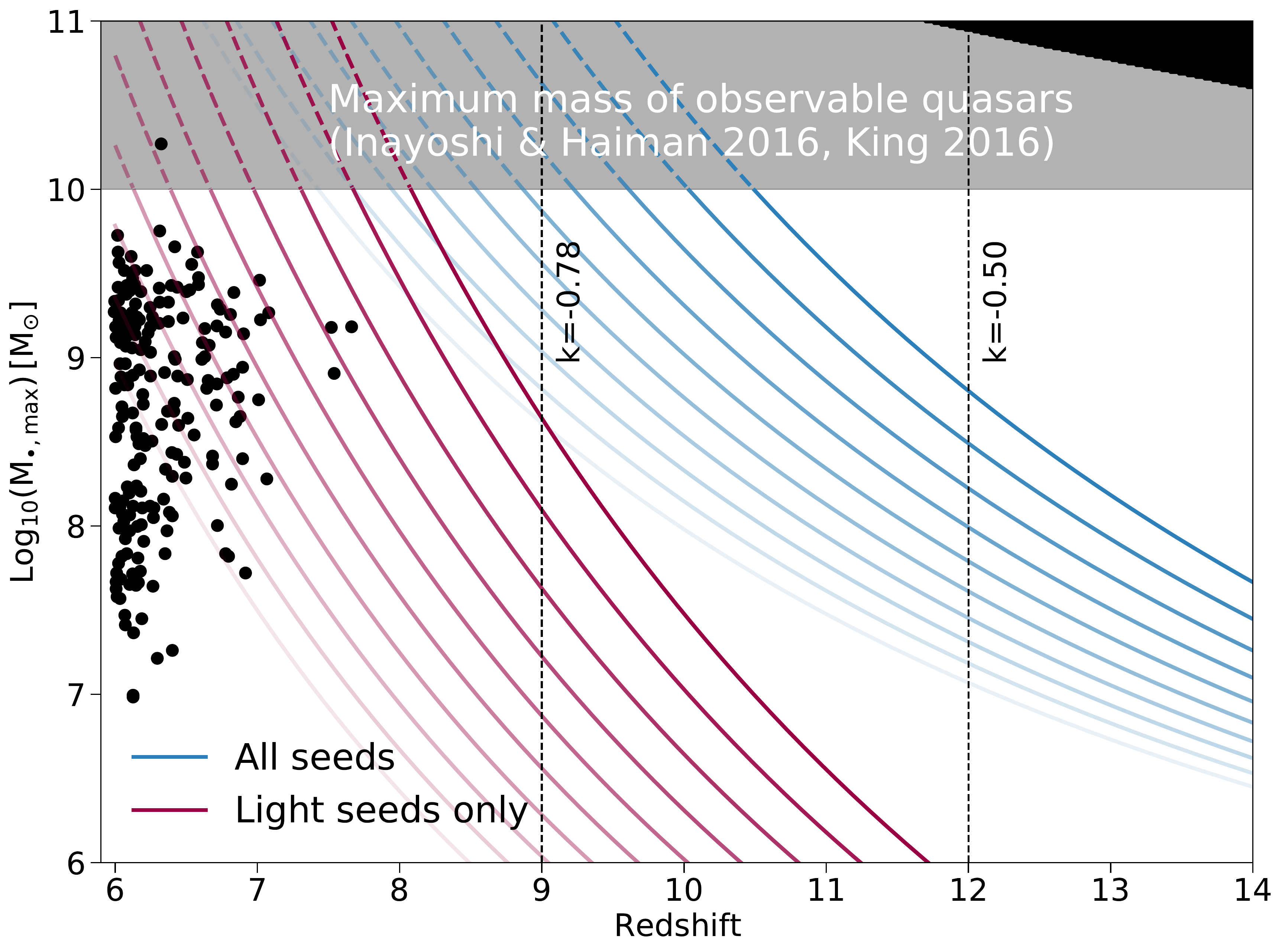}
\caption{Maximum mass achievable by a black hole seeded at $z=25$ which undergoes continuous Eddington-limited accretion. Red lines: the seed mass is within the light seeds regime, i.e. $\sim 10^2 \Msun$. Blue lines: the seed population also includes heavy seeds, with mass $\sim 10^5 \Msun$. The different levels of transparency show the radiative efficiency assumed, from $\epsilon=0.057$ (no transparency) to $\epsilon=0.1$ (maximum transparency). The black circles indicate the $\sim 200$ quasars at $z > 6$ known so far. The vertical, dashed lines show the current redshift range for the farthest $\Mblack \gtrsim 10^9 \Msun$ quasar observable by future surveys, while the shaded regions on top show the current limits for the maximum mass of a black hole observable (see text for a detailed description).} 
\label{fig:max_mass}
\end{figure}

In Fig. \ref{fig:max_mass} we show the maximum mass achievable by a black hole seeded at $z=25$, assuming continuous (${\cal D}=1$) Eddington-limited ($\fedd=1$) accretion. We consider two seeding scenarios: (i) ``All seeds'' includes light and heavy seeds up to a typical mass $10^5 \Msun$, while (ii) ``Light seeds only'' includes only light seeds up to a typical mass $10^2 \Msun$. The growth lines sharing the same color but different transparency levels describe accretion scenarios with varying radiative efficiencies, from $\epsilon=0.057$ (no transparency) up to $\epsilon=0.1$ (maximum transparency), in ten steps. Even a change by a factor $<2$ in the radiative efficiency can have dramatic consequences on the predicted mass growth, up to $\sim 3$ orders of magnitude by detection redshift $z_{\rm det}=9$. A better determination of the probability distribution of $\epsilon$, which will likely come from higher redshift detections (see \S \ref{sec:parameters}), is of primary importance.

Some of the current $z > 6$ detections of quasars, shown as black circles in Fig. \ref{fig:max_mass}, already rule out several models of growth. For example, a light seed mass $\sim 10^2 \Msun$ with continuous Eddington accretion is currently ruled out for any $\epsilon > 0.07$. Note that, using the dimensionless spin parameter $0 < a < 1$ ($a = 0$ for a non-rotating black hole and $a = 1$ for a maximally rotating black hole), the previous lower limit sets a limit on the typical SMBH spin for thin-disk accretion: $a > 0.3$.

Data for the $\sim 200$ detected quasars at $z > 6$ is obtained from \cite{Inayoshi_review_2019} and integrated with recent detections, i.e., \cite{Yang_2020, Wang_2021}. The light shaded region indicates the approximate maximum mass observable in quasars that are radiating energy via their accretion disk \citep{King_2016_MaxMass, Inayoshi_2016_MaxMass}, while the black shaded region indicates an absolute upper limit on the mass reachable by an accreting black hole at a given redshift. It is obtained by calculating at each $z$ the maximum mass of a dark matter halo whose number density is equal to $1/V(z)$, with $V(z)$ being the comoving volume of the Universe at that redshift. We assume a Sheth-Mo-Tormen \citep{SMT_2001} halo mass function and a full-sky survey with $4\pi$ steradians of coverage. This mass is then converted into baryons assuming a baryonic fraction $f_b = 0.15$. The maximum mass is obtained assuming that the black hole is able to accrete the baryonic mass of its host in its entirety, in what is a very solid upper limit. 

The two vertical, dashed lines show the current predictions \citep{Fan_2019BAAS, Wang_2019_density, Euclid_2019} for the farthest $\Mblack \gtrsim 10^9 \Msun$ quasar observable assuming a rate of quasar volume density decrease with redshift $k$, i.e. $\rho_{\rm QSO} = 10^{-kz}$.
These limits show any detection of $z>9$ quasars in future surveys will provide crucial insights in shrinking the parameter space for growth parameters available. In particular, a detection of a quasar with $\Mblack \sim 10^{10} \Msun$ by $z\sim 9-10$ would exclude entirely the parameter space available for light seeds and dramatically reduce the one for massive seeds (unless we account for continuous accretion in highly radiative inefficient scenarios, i.e. $\epsilon < 0.057$, see \S \ref{sec:disc_concl}).

It is worth noting that future high-$z$ quasar surveys (e.g., Euclid and the Large Synoptic Survey Telescope, or LSST) may have the sensitivity to detect quasars with masses down to $10^{7-8} \Msun$, further constraining their demographics. Nonetheless, extreme SMBHs with masses $\gtrsim 10^9 \Msun$ undoubtedly play a more significant role in shrinking the parameter space available for the growth parameters. In fact, statistical distributions, such as those in Fig. \ref{fig:Cornerplot}, obtained for the formation of a $10^8 \Msun$ and a $10^7 \Msun$ SMBH by the same redshift have typical standard deviations $\sim 35\%$ and $\sim 70\%$ larger, respectively. For example, the detection of a $\Mblack \sim 10^{7-8} \Msun$ quasar by $z \sim 9$ would still broadly allow a seed mass of $M_{\rm \bullet, seed} \sim 10^3 \Msun$ and Eddington ratios and duty cycles as low as $\sim 0.5$. On the contrary, the detection of the same lower-mass quasars at $z\sim 12$ would more significantly constrain the parameter space, with a seed mass $M_{\rm \bullet, seed} \gtrsim 10^{4.5}$ now strongly favored.

\section{Discussion and Conclusions} 
\label{sec:disc_concl}
The highest-redshift quasar currently stands at $z=7.642$, with a typical mass $\sim 10^9 \Msun$ \citep{Wang_2021}. In this paper we studied how the detection of higher-redshift quasars, which will most likely occur with upcoming surveys (e.g., \citealt{Fan_2019BAAS, Wang_2019_density, Euclid_2019}), will shed light on the growth parameters involved in high-$z$ black hole growth, as well as enhancing our knowledge on the mass distribution of seeds. 
We quantified how, as the time between seeding and observation is reduced with increasing detection redshift, the combination of growth parameters that can give rise to a specific black hole mass shrinks significantly.
Hence, upcoming $z>7$ quasars surveys will play a crucial role in the determination of the allowable parameter space for black hole growth.

It is important to point out that the broad range of our flat priors are not due to lack of complexity in our model, but reflect the real uncertainties in this field. The primary goal of our study is to show how these uncertainties will be reduced by higher-redshift detections.

Obviously, there are caveats and limitations to our study. 
First, we assume that the growth parameters are constant from seeding to detection redshift. As discussed in \S \ref{subsec:MC}, this is equivalent to consider them as averages of variable parameters over the growth time. While a reasonable assumption, the average is not representative of extreme values which can occur in the parameter space and that can speed up or hamper the growth. A distribution of randomly changing parameters, possibly with a frequency matched to some fraction of the Eddington time, would be more realistic, but any choice of this type would be even more arbitrary than considering time averages. For this reason, we stand by our choice to consider averages of $\left[ \fedd, {\cal D}, \epsilon, M_{\rm \bullet, seed} \right]$ as a good representation of the growth process over timescales comparable to the Eddington time.
Second, in Fig. \ref{fig:max_mass} we showed how the maximum mass achievable depends strongly on the radiative efficiency of the accretion process. For example, a mere change in the radiative efficiency parameter $\epsilon$ by a factor $2$ could change the maximum mass achievable by $\sim 3$ orders of magnitude by $z\sim 9$ in the light seeds only scenario. In this study we considered only a standard thin disk accretion, with values for $\epsilon$ ranging between 0.057 and 0.32 \citep{Bardeen_1970, Fabian_2019}. Alternative scenarios have been proposed in highly inefficient accretion flows, where the radiative efficiency is even lower. For example, in the slim disk scenario \citep{Mineshige_2000, Volonteri_2014, PVF_2015} a typical radiative efficiency of $\epsilon = 0.04$ is achieved for black holes accreting at ${f_{\rm edd}} \sim 50$. With a constant Eddington-limited accretion from $z=25$, a light black hole seed with such a radiative efficiency would grow to a mass in excess of $10^{10} \Msun$ by the detection redshift $z=9$.

For all these reasons, investigating the typical radiative efficiency of high-$z$ accretion lays at the basis of any of our attempts to constrain growth parameters and seeding models for black holes.
While large quantities of gas in the high-$z$ Universe would favor inefficient accretion flows and lower values of the radiative efficiency (e.g., \citealt{Begelman_Volonteri_2017}), a stable accretion via a disk would spin up the black hole (e.g., \citealt{Berti_Volonteri_2008}), leading to an increase of the radiative efficiency and hampering the growth \citep{Novikov_Thorne}. The interplay between these two factors in shaping the growth history of early black holes remains to be seen.

Future quasar surveys will be fundamental to study the parameter space available for growth and seeding models. Gravitational lensing, by magnifying the light of extremely faint sources above the flux limit, will enhance our chances of detecting very early SMBHs \citep{Fan_2019, Pacucci_Loeb_2019}. While the predicted yields of quasar surveys suggest that the farthest detectable quasar with $\Mblack \gtrsim 10^9 \Msun$ will be somewhere between $z=9$ and $z=12$, these predictions are based on extrapolations and may reserve surprises. What if the spatial density of quasars decreases much more rapidly than predicted, and we observe the highest-redshift quasar by $z\sim 8$? This would push to relaxing some of the requirements for building high-$z$ black holes, leaving much of the parameter space still available. Even more interestingly, what if the spatial density decreases much more slowly, or even reaches a plateau? This instance would require a full review of the bottom-up paradigm of black hole growth, seeking alternative models that are able to create already very massive black holes at high-$z$ (e.g., primordial black holes, \citealt{Carr_2018}).

Obscured accretion phases in $z\gtrsim 7$ quasars may significantly impede (or bias) their observations. For example, \cite{Ni_2020}, using the \textsc{BlueTides} cosmological simulations, predict that large fractions (60\% to 100\%) of $z > 7$ quasars with X-ray luminosity $L_X > 10^{43} \, \mathrm{erg \, s^{-1}}$ may be heavily obscured (i.e., with a column density in excess of $10^{23} \, \mathrm{cm^{-2}}$). Additionally, they find that $>99\%$ of $z>7$ quasars are dust extincted and, hence, non detectable in ultraviolet surveys. If obscuration indeed reduces the yields of future surveys, it may also push the ``farthest detectable quasar'' to lower redshifts. In this case, growth models will need to rely on gravitational waves observations to detect higher-redshifts SMBHs \citep{Ricarte_2018, Pacucci_2020}.

In one way or another, future quasar surveys will inevitably push the boundaries of the unknown and reveal some of the mysteries that still surround the growth and early years of these cosmic giants.

\section*{Acknowledgements}
The authors thank Xiaohui Fan for insightful comments on the manuscript, as well as inspiring discussions with Tyrone Woods. We also thank the anonymous referee for constructive comments on the manuscript.
F.P. acknowledges support from a Clay Fellowship administered by the Smithsonian Astrophysical Observatory. This work was partly performed at the Aspen Center for Physics, which is supported by National Science Foundation grant PHY-1607611. The participation of F.P. at the Aspen Center for Physics was supported by the Simons Foundation.
This work was also supported by the Black Hole Initiative at Harvard University, which is funded by grants from the John Templeton Foundation and the Gordon and Betty Moore Foundation.

\section*{Data Availability}
The data underlying this article will be shared on reasonable request to the corresponding author.



\bibliographystyle{mnras}
\bibliography{ms}




\bsp	
\label{lastpage}
\end{document}